# A MOLECULAR MASS GRADIENT IS THE KEY PARAMETER OF THE GENETC CODE ORGANIZATION

Felix Filatov

*Hematology Research Center*
*Russian Academy of medical sciences, Moscow, Russia*
(e-mail ffelix001@yahoo.com, ffelix@blood.ru)

**ABSTRACT**

The structure of the genetic code is discussed in formal terms. A rectangular table of the code ("the code matrix"), whose properties reveal its arithmetical content tagged with the *information symbols* in several notations. New parameters used to analyze of the code matrix, the serial numbers of the encoded products and coding elements, ordered by molecular mass. The structural similarity of the amino acid sequences corresponding to two aminoacyl *tRNA* synthetases classes is found. The code matrix shows how can be organized the so-called *second genetic code*. The symmetrical pattern of the matrix is supported with the other parameters; it also serves as a basis to construct a 3D model of the genetic code which follows the structure of the simplest Plato solid, tetrahedron. The reasons for this unusual structure of the genetic code remains unclear.

## 1 INTODUCTION

The genetic code (a set of the instructions to convert a nucleotide sequence into a polypeptide one) is the central phenomenon connecting the world of nucleic acids and the world of proteins that marked the start of biogene evolution [01]. 2-3 billion years preceed this event was supposed to be enough to synthesize nitrogen bases and amino acids, and also to form the machinery of their polymerization. The emergence of adapter molecules, *tRNA*, capable to bind a matrix polymer and a growing polymer chain of amino acids became a basic prerequisite for the formation of the code. This binding effectively realizes only when *tRNA* is acetylated with special enzyme, aminoacyl-*tRNA* synthetase, *AARS*. So, several molecular components are involved in a single act of the genetic coding. Three of them use non-complementary mutually recognition mechanisms: (1) *tRNA* with its anticodon corresponding to a specific amino acid, (2) the amino acid itself, and (3) *AARS* corresponding to a specific *tRNA* and a specific amino acid. The problems are how the *AARS* molecule recognizes the *tRNA* corresponding to a target amino acid, and what the origin of this correspondence is. All *AARS* are divided into two classes based on several criteria (specificity of an ATP site, fixation on a small or a large groove of the *tRNA* acceptor stem, etc.). We are not discussing here a few non-significant violations of this division. At the same time the variability of *tRNA* acceptor stem is not thus much to form the so-called *second genetic code* similar to the *first one* by a structure [02]. Recognition of the *tRNA* anticodon by the *AARS* molecule is rather the conformance exception than the rule [03]. Significant regularity of the genetic code allows building its various symmetric models and generates assumptions of the possible predecessors of the coding machinery [05-10], or of the principal limitations of the molecular biology axiomatics as basis for hypotheses of the origin of the genetic code [04].

## 2 PROBLEM

We believe that a key factor in the organization of the genetic code should be a sensitivity of *AARS* molecules to the gradient of the molecular masses of the *tRNA* anticodon region bases (without production of a stable bond) rather than to the other structural areas of the *tRNA* molecule. To differentiate between two chemically similar doublets (e.g. *CA* and *AC*) *AARS* molecule has to recognize "its own" side of the acceptor stem of *tRNA* (large or small groove) and roughly estimate bulkiness of the 2$^{nd}$ and the 3$^{rd}$ anticodon bases corresponding to a codon doublet (1st and 2nd bases) and their mutual orientation. To complete the recognition molecule the *tRNA* has to have additional specific features on its surface. The mutual recognition *tRNA* and *AARS* follows to the so-called *second genetic code* which has to correspond to the first one precisely. *The first genetic code* has clear marks of all this following two rules (*Table* 1):

1. *heterodoublet of nitrogenous bases with increasing molecular masses encodes a heavier amino acid than heterodimer of the same bases with decreasing masses* (exception - *AG* [**S**, **R**]; hereinafter we used one-letter symbols for purines, pyrimidines and amino acids;
2. *homodoublet of two same pyrimidines (true homodoublet) encodes heaver amino acids than homodimer* of two same purines (exception - *TT* [**L**]).

***Table* 1.** Dependence of the molecular mass of the coded amino acids on the gradient of the coding doublet nitrogen bases molecular mass.

| YY | RY | | | | RR | RR | RR |
|---|---|---|---|---|---|---|---|
| TC | AC | GC | AT | GT | GA | GG | AA |
| S | T | A | I M | V | D E | G | N K |
| L | Q H | R | Y | C W | S R | P | L F |
| CT | CA | CG | TA | TG | AG | CC | TT |
| YY | yR | | | | RR | YY | YY |

Left - *heterodoublets*. The increasing or the decreasing molecular mass are shown with size of the nitrogen bases symbols. The top and the bottom lines show the purine-pyrimidine symbols of the coding doublets content. Pairs purines and pair pyrimidines marked with gray are *conditional homodoublets*. Black symbols on light gray – the amino acids of less molecular mass; white symbols on dark gray – the heavier amino acids. The latter contains three exceptions of two rules described in the text (black symbols without a background); in some minor codes the triplet *AGR* corresponding the "universal" **R** encodes nothing.
Right - *True homodoublets* (no mass gradient; see the text).

Heterodoublets of two different purines or two different pyrimidines (*conditional homodoublets*) follow the first rule (the rule of priority). The proximity of molecular masses of *C* and *U* is probably resulted synonymous doublet reading *CU* and *UU* as the same amino acid (**L**). Curiously, the same antidoublet (anticodon doublet complementary to codon doublet) *UC* encodes amino acids **S** and **R**, which are also coded by different doublets *UC* (**S**) and *CG* (**R**).

We tried to use the same formalization (ordering of amino acids by molecular mass) to understand the organization principles of the genetic code. The attempts to formalize of the code repeatedly described elsewhere in terms of algebra, geometry, topology, theory of information *etc* [05-10] show the impossibility to understand the code origin only in the frames of the molecular biology axiomatics. Based on algebra *Yang* [11] constructed the sophisticated 28-gone 3D-model of the genetic code and found some association of his model with division of the *AARS* into two classes. *Negadi* [12] has shown that the *Yang's* model [11] associates with *shCherbak's* concept of the arithmetic content in the genetic code [09]. *Swanson* [08] and *Bosnacky et al* [13] showed that organization of the genetic code has obvious similarities with the Gray code. *Patel* found that the choice of the number 20 (number of amino acids) corresponds to solutions of some equations of the theory of information [07]. Nevertheless an exclusively simple version of the formalization of the genetic code described here seems to us worthy to discuss.

The classification of molecules participating in polypeptide synthesis divides them into subgroups allowing a much more structured presentation of the genetic code than used in textbooks. One of these groups is the four nitrogenous bases which can be divided either into two purines and two pyrimidines *CT*(*YY*) and *AG*(*RR*) or into two complementary pairs *GC* and *AT*. Another group of molecules of the genetic coding is the aminoacyl *tRNA* synthetases, *AARS*, which also can be divided into two equal (10 and 10) classes [14]. Since the accuracy of the mutual recognition of codon-anticodon (*mRNA-tRNA*) pair should be comparable with the accuracy of recognition of corresponding pairs of *AARS-tRNA* and *AARS*-(*amino acid*) we believe useful to divide these molecules (20 *tRNAs* and 20 *amino acids*) into two corresponding classes too, *aars* 1 (*Y*) and *aars* 2 (*R*), see below.

## 3  RESULTS

### Structuredness of the ordered sequences of the coded products

We ordered two classes of amino acids corresponding to the 1$^{st}$ and the 2$^{nd}$ *AARS* classes (excluding three above mentioned amino acids **S**, **L** and **R** encoded by the *Rumer octet 2* doublets [15]), *Table* 2.

*Table 2*. Ordering of the amino acids of two *aars* classes, *Y* (**A**) and *R* (**B**), by molecular mass.

**A**

| *arcc* *Y* | V | C | L | I | Q | E | M | R | Y | W | AA |
|---|---|---|---|---|---|---|---|---|---|---|---|
| | G | T | C | A | C | G | A | C | T | T | 1$^{st}$ ltr |
| | 1 | 2 | 3 | 4 | 5 | 6 | 7 | 8 | 9 | 10 | 1,2,3… |
| | 43 | 47 | 57 | 57 | 72 | 73 | 75 | 100 | 107 | 130 | NM |

**B**

| *arcc* *R* | G | A | S | P | T | N | D | K | H | F | AA |
|---|---|---|---|---|---|---|---|---|---|---|---|
| | G | G | T | C | A | A | G | A | C | T | 1$^{st}$ ltr |
| | 1 | 2 | 3 | 4 | 5 | 6 | 7 | 8 | 9 | 10 | …1 |
| | 10 | 9 | 8 | 7 | 6 | 5 | 4 | 3 | 2 | 1 | …2 |
| | 11 | 12 | 13 | 14 | 15 | 16 | 17 | 18 | 19 | 20 | …3 |
| | 1 | 15 | 31 | 41 | 45 | 58 | 59 | 72 | 81 | 91 | NM |

*The top lines* – the amino acid (*AA*) symbols; intensity of gray follows the increasing of the amino acids molecular mass expressed in terms of the *nucleon mass* (**NM**) in *the bottom lines*.
*The second lines* – the 1$^{st}$ letters (*1$^{st}$ ltr*) of the codons corresponding the amino acids above. Repeated blocks are marked with black background; unique pairs (*CT* in the *aars* class *Y*, and *GA* in the *aars* class *R*) are marked with dark gray.
*The rest lines* – direct version of separate numbering of the amino acids (#**1**: 1-10-1-10 of each *aars* class) and reverse one (#**2**: 1-10-10-1), and also general numbering (version #**3**: 1-20). Intensity of gray follows the increasing of the amino acids numbers.

The order of the amino acids of equal molecular mass is determined by the mass of the first letter of their coding triplets. Both sequences turn out to be clearly structured by the 1$^{st}$ coding triplet letter. Each class contains two blocks of four amino acids (1$^{st}$ codon letters *GTCA* and *GATC*) and also two pyrimidines (the *aars Y* corresponding to the *AARS*-1) and two purines (the *aars R* corresponding to the *AARS*-2).

### The matrix of the genetic code and the position symmetries of the coded products

Placing 20 amino acids in a rectangular table 4x5 we obtain what we called *the matrix of genetic code* (*Table* 3). The vertical coordinate of the matrix is formed by the 1$^{st}$ codon letters ordered by mass, the horizontal one is formed by corresponding amino acids also ordered by mass. The matrix has several curious properties. One of them is the symmetry of the distribution of the amount of amino acids of each *aars* class in the columns and in the mutually "complementary lines" (about the *Y|R* axis separating the first encoding pyrimidines and purines). The matrix allows organising an additional (central) line containing three amino acids and two punctuation signals that are not included in the basic matrix version.

*Table* 3. The matrix of the genetic code and its symmetries.

| | | 1 | 2 | 3 | 4 | 5 | 1 | 2 | 3 | 4 | 5 | aarsR : aarsY |
|---|---|---|---|---|---|---|---|---|---|---|---|---|
| Y | C 35 | P | L | Q | H+ | R+ | P | L | Q | H | R | 2:3 |
| Y | T 50 | S | C | F | Y | W | S | TG | TT | TA | W | 2:3 |
| | | | | | | | S | 0 | L | 0 | R | |
| R | A 59 | T | I | N | K+ | M | T | I | N | K | M | 3:2 |
| R | G 75 | G | A | V | D- | E- | G | A | V | D | E | 3:2 |
| aarsR : aarsY | | 4:0 | 3:1 | 2:2 | 1:3 | 0:4 | | | | | | |

<u>Left</u> – "*charged version*" of the code matrix. The columns are marked with digits *1-5*, the lines are marked with the 1st codon nitrogen base symbols of the order *C<T<A<G*. The gray spheres - the amino acids of the ***aars Y*** class (light gray - the *invariant monomers* **L** and **Y**, see text); the light spheres – the amino acids of the ***aars R*** class (black symbol in gray - the *invariant monomer* **K**, see text). The number below a symbol of the nitrogen base – its nucleon mass. Ratio ***aars R***: ***aars Y*** by symmetrical columns (1|5, 2|4) and by symmetrical (complementary - in terms of the 1st codon letters: *CG=AT*) lines is marked with gray and white. The central column (*3*) is highlighted with gray.

<u>Right</u> - additional (central) line *AT* above the matrix plane (gray spheres); it consists of three amino acids coded with two octet **2** doublets, and two stop codons (*0*). Products coded by triplets with the 1st *A* (**S,R**) are located in the same columns as same "conventional" products; products coded by triplets with the 1st *T* (*0*,**L**,*0*) are located in the same columns as corresponding doublets (*TG,TT,TA*).

Another group of symmetries is, *e.g.* the symmetry of so called the *degeneracy groups* about the axis *Y|R*. The genetic code is degenerated: each of 8 amino acids of the degeneracy group **IV** encoded by **four** triplets, each of 2 amino acids of the degeneracy group **III** encoded by **three** triplets, each of 12 amino acids of the degeneracy group **II** encoded by **two** triplets, and only 2 amino acids of the degeneracy group **I** encoded by **one** triplet. The first 8 amino acids of the degeneracy group **IV** form the Rumer *octet 1*, the rest 15 amino acids (including repeated **S**, **L** and **R**) of the degeneracy groups **III**, **II** and **I** form the Rumer *octet 2* [15]. The "complementary lines" of the matrix (*C* and *G*, *A* and *T*) contain an equal amount of amino acids of each degeneracy group. As a rule the number of the degeneracy group decreases along the line - from left to right.

**The matrix of the genetic code and the numeric symmetry of the coded products**

The matrix positional symmetry (the symmetrical pattern) shown above is combined well with the symmetries determined by the positions of amino acids in ordered sequences. Numbering of amino acids in the sequence of each of the two ***aars*** classes creates a new (*virtual*, non-physical) integer parameter of their molecules. Three variants of this numbering depending of different directions of each class, (1) 1-20, (2) 1-10-1-10, (3) 1-10-10-1, are shown in *Table* 2. Numbering of amino acids according to their hydrophobicity was first used by *Lacey & Mullins* [16], but their sequences did not demonstrate symmetric patterns like those described here. Analysis of sums of these parameters in lines and columns of the genetic code matrix shows strict numerical symmetries (*Table* 4). The matrix pattern in all three variants highlights three central columns that suggests non-random organization of the genetic code and its matrix, and both pairs of the "complementary" lines. This pattern is marked with specific three-digit numbers ($n111_b$, where $n$ is integers 1, 2 or 3, and $b$ is the base of quaternary, quinary or decimal notations) called *the information symbols* [09].

*Table 4.* The matrix of the genetic code corresponding to three versions of the enumeration of the amino acids

**A.**

|   | 1st base |   | 1 | 2 | 3 | 4 | 5 |   |
|---|---|---|---|---|---|---|---|---|
|   | C | 1 | **04** | **03** | **05** | **09** | **08** | 29 |
| Y | T | 2 | **03** | **02** | **10** | **09** | **10** | **34** |
| R | A | 3 | **05** | **04** | **06** | **08** | **07** | 30 |
|   | G | 4 | **01** | **02** | **01** | **07** | **06** | 17 |
|   |   |   |   |   |   |   |   | 110 |
|   | *Decimal notation* |   | *13* | *11* | *22* | *33* | *31* |   |

**B.**

|   | 1st base |   | 1 | 2 | 3 | 4 | 5 |   |
|---|---|---|---|---|---|---|---|---|
|   | C | 1 | **07** | **03** | **05** | **02** | **08** | 25 |
| Y | T | 2 | **08** | **02** | **01** | **09** | **10** | **30** |
| R | A | 3 | **06** | **04** | **05** | **03** | **07** | **25** |
|   | G | 4 | **10** | **09** | **01** | **04** | **06** | 30 |
|   | *Decimal notation* |   | **31** | 18 | 12 | 18 | **31** | 110 |
|   | *Quinary notation* |   | *111₅* | *33₅* | *22₅* | *33₅* | *111₅* |   |

**C.**

|   | 1st base |   | 1 | 2 | 3 | 4 | 5 |   |
|---|---|---|---|---|---|---|---|---|
|   | C | 1 | **14** | **03** | **05** | **19** | **08** | 49 |
| Y | T | 2 | **13** | **02** | **20** | **09** | **10** |   |
| YR | TA | 2+3 | **13** | *0* | **03** | *0* | **08** | **69x2** |
| R | A | 3 | **15** | **04** | **16** | **18** | **07** |   |
|   | G | 4 | **11** | **12** | **01** | **17** | **06** | 47 |
|   | *Decimal notation* |   | 53 | 21 | 42 | 63 | 31 | 210 |
|   | *Quaternary notation* |   | **311₄** | *111₄* | *222₄* | *333₄* | *133₄* |   |

Summations of the amino acids position numbers by columns and lines are shown under corresponding columns and right of corresponding lines. Black numbers in white are symmetrical (by positions of digits) to white numbers in black (shown with symbols "~" or "=").

**A.** The numbering version **1**. Lines $C+G \sim A+T$: [30+34=**64**]~[29+17=**46**], columns 1 and 5: **13~31**.

**B.** The numbering version **2**. Lines $C+G \sim A+T$: **55=55**, columns 1 and 5, 2 and 4: **31=31**, 18=18.

**C.** The numbering version **3**. Lines $C+G \sim A+T+TA$: [after insertion the *additional line*, *Table 3*]: **69**(x2)~**96**, columns 1 and 5: **311~133**.

Increasing the molecular mass of the genetic code components is marked with intensity of gray. General sums of numbers $2_x(1+2+\ldots9+10)=110$ corresponding to the versions **1** and **2**, and $(1+2+\ldots+19+20)=210$ corresponding to the version **3** are in gray plaques.

**The matrix of the genetic code and the quantitative symmetries of the coded products**

These symbols illustrate the arithmetic structure of the genetic code which can also be seen in terms of so called *nucleon mass* of the encoded products. The *nucleon mass* (parameter introduced for the first time by *Hasegawa and Miyata* [17]) represents the number of nucleons (protons and neutrons) of atoms of amino acids molecules. In this study we expand the concept of the *nucleon mass* to the other molecular constituent of the genetic code, nitrogen bases. Following the approach of Hasegawa & Miyata who divided amino acids into constant and variable (side chain) parts we also divide the nitrogen bases into constant (their hexacycle) and variable (combination of all atoms outside the hexacycle) parts. The nucleon mass of the hexacycle is 76, and the nucleon mass of variable part is 35 for *C*, 50 for *T* (38 for *U*), 59 for *A*, and 75 for *G*. One can see that nucleon sum of the complementary pairs are equal: *C+G* (minus 3 protons of the hydrogen bond) = *A+T* (minus 2 protons of the hydrogen bond) = 107. This balance marks the DNA stability. The RNA with its *C<U<A<G* set of the nitrogen bases lacks this balance, and perhaps this indicates that the genetic code has completed its evolution (or at least its stability) when the RNA world was supplemented by a DNA.

The matrix of the genetic code demonstrates the quantitative orderliness and symmetries about the vertical (central column) and horizontal *Y|R* axes - in terms of the nucleon mass of the nitrogen bases and amino acids. The first group of symmetries refers to the nucleon masses summations in the columns of the *charged variant* of the matrix (*Table* 5). In several notations these balances are also marked by the *information symbols*.

*Table* **5**. The matrix of the genetic code (*the nucleon* variant) and its symmetries.

| | | 1 | 2 | 3 | 4 | 5 | 1 | 2 | 3 | 4 | 5 |
|---|---|---|---|---|---|---|---|---|---|---|---|
| Y | $C_{35}$ | IV | IV | II | II | IV | 41 | 57 | 72 | 82 | 101 |
| | $T_{50}$ | IV | III | II | II | I | 31 | 47 | 91 | 107 | 130 |
| R | $A_{59}$ | IV | III | II | II | I | 45 | 57 | 58 | 73 | 75 |
| | $G_{75}$ | IV | IV | IV | II | II | 01 | 15 | 43 | 58 | 72 |
| | | | | | | | | $[2 \times 444 + 4^2]_5$ | | | |
| | | | | | | | | $[4 \times 444]_5$ | | | |
| | | | | | | | | $[4 \times 444]_5$ | | | |

<u>Left</u> – Symmetry (about the central horizontal axe – *Y|R*) of the amino acids by degeneracy groups (see text). White symbols on black – *the degeneracy group* **IV** (the Rumer *octet 1*), black symbols the Rumer *octet 2*, the degeneracy groups **III** (dark gray), **II** (light gray), **I** (transparent). Quantities of amino acids of the same degeneration group are equal in "complementary lines".

<u>Right</u> – Symmetry (about the central vertical axe – column *3*) of the column sums of the amino acids nucleon masses expressed with the *information symbols* in quinary notation ("*charged version*" of the code matrix). In the "*neutral version*" - $H_{81}$, $R_{100}$, $D_{59}$, $E_{73}$, $K_{72}$ (the amino acid nucleon mass is right of its symbol) – ratio of the total nucleon mass of the "pyrimidine (*Y*) lines" (757) and the total nucleon mass of the "purine (*R*) lines" (498) is 3/2 (curiously, ratio *Y* and *R* is approximately inverse: ~2/3). Three "extra" amino acids, $S_{31}$, $R_{100}$ (lines *R*) и $L_{57}$ (lines *Y*), make this ratio remarkable in decimal notation

$$Y:R = (777+q) : (555+2q),$$

where *q* = 37 (*111*:3) is the so-called *prime number of the genetic code* [09]. The values of the nucleon mass of six amino acids corresponding to the *information symbols* in several notations with bases 5, 6, 7, 8, 9 (31=$111_5$, 43=$111_6$, 57=$111_7$, 72=$111_8$, 91=$111_9$) are highlighted with white on black plaques. Total sums of these bases in the *Y* lines and *R* lines are equal: 5+7+9 = 6+7+8 = 21 = $111_4$.

The matrix of genetic code allows constructing unique 3D model based on the simplest Plato solid, tetrahedron of 20 equal spheres (monomers) with quantitative symmetry of its faces (in terms of sum of the cumulative *nucleon mass* of 10 amino acids of each face). These monomers can be divided into two groups. One of them is composed of four vertices and four face centers of tetrahedron. Both quartets can be exchanged preserving the values of *the nucleon* *mass* of faces; we call them here *invariant* monomers. The other group consists of six pairs of inner monomers of six edges (*edge* monomers) of the tetrahedron. If each monomer corresponds to *its own* amino acid of the certain *nucleon mass* then the symmetric position of the *edge* and the *invariant* monomers about *Y|R* axe allow constructing the tetrahedron of bilateral quantitative symmetry (the mass of its two faces is equal to the mass of the other two faces), *Table 6*. Each face of the tetrahedron contains the same numbers of amino acids of both *aars* classes: 5+5.

*Table 6*. The matrix of the genetic code (*the nucleon* variant) and its 3D (tetrahedron) model.

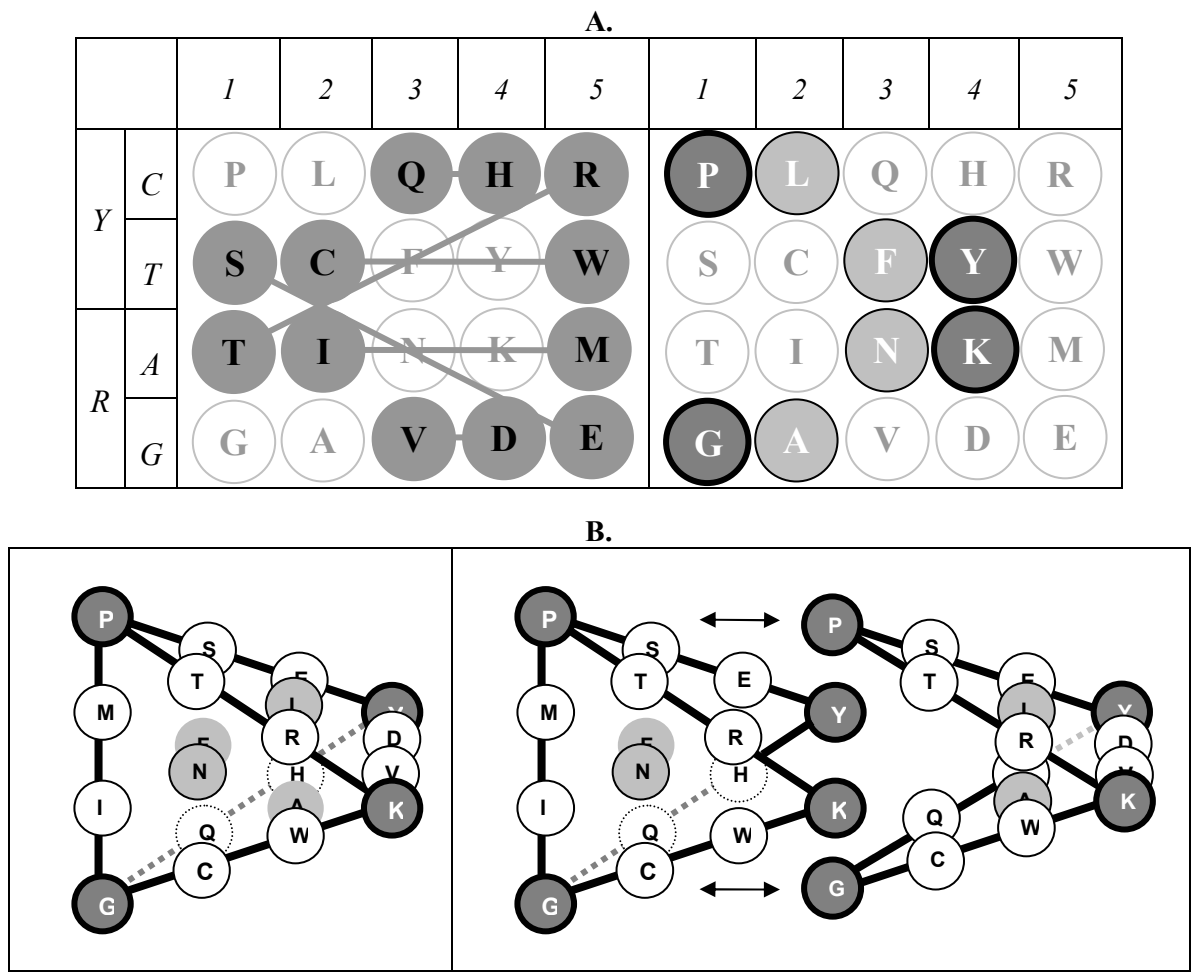

**A.** Left - positions of six pairs of the *inner edges monomers* of the code tetrahedron (gray circles; see text) connected with straight lines (the *neutral version* of the matrix).
Right - positions of the *invariant monomers* of the code tetrahedron (dark grey circles - vertices, light grey circles – faces centers; see text).
**B.** Left – the tetrahedron (3D model) of the genetic code.
Right - its bilateral symmetry: total nucleon mass of each two faces is 1255 (626+629=627+628).

## 4 DISCUSSION

The regularity of the sequence of the amino acids ordered by mass and divided into two *aars* classes (corresponding to two *AARS* classes) indicates a key role of this classification in the organization of the code. The main feature of this regularity, the sequence of the 1$^{st}$ codon letters, allows using it as the second parameter of a geometric regular table called here *the matrix of the genetic code*. All data presented here show that its pattern can not be just accidental. Perhaps the matrix can provide a key to understanding how the *AARS* recognize the third component of the coding system, *tRNA*. Structural basis for the first step of this recognition which reproduces the classification of *AARS* is two opposite grooves of the acceptor stem of *tRNA* molecule. This, however, is not sufficient for unerring compliance of each (of 20) *AARS* to each (of 20) group of the isoacceptor *tRNA*s containing anticodons for each (of 20) amino acids encoded this way. This compliance (or so-called *the 2nd genetic code*) requires that each *tRNA* molecule should have one of 20 well distinguishable and recognizable by *AARS* sites. Despite intensive searches, such sites have not yet been identified. And yet the matrix of the genetic code well illustrates that the amount of these sites can be not this much. If the *AARS* molecule can recognize the anticodon 3`-base (corresponding to the codon 5`-base) by comparing its size with the downstream methylated purine (inosine) size, then only two or three specific sites would be enough to the final recognizing of the corresponding *tRNA* (two or three are the numbers of amino acids of each *aars* class in each of the matrix line).

Suggested mechanism of recognizing *tRNA*s by *AARS* molecules does not explain, however, why the pattern of the genetic code shown here is that it is. Attempts to answer the question *does the modern table of the genetic code contain any hint of its origin* lead usually only to repeated appearance of new and new tables of the code. One of these tables is discussed in this paper. Its basic feature is the development and deepening of the shCherbak thought of the arithmetic content of the genetic code. The arithmetic of the genetic code can be demonstrated not only with decimal, as shCherbak has described [09], but also with other notations as it has been shown here.

The organization of the matrix of the genetic code is based on a gradient of molecular mass of neighbor amino acids and nitrogen bases in the linear representations. This gradient determines the order of amino acids or nitrogen bases and in the latter case it also indicates the comparability of the ordering by molecular mass with complementarity. Two **aars** classes of amino acids show the same order of corresponding $1^{st}$ codon letters in the linear and in the matrix presentations. Rather it indicates a parallel processes of the forming the code for amino acids of both classes. At the first glance the second triplet bases are disordered, and it probably indicates the evolutionary priority of molecular mass over the other physical properties of the code products, *e.g.* hydrophobicity, which is associated with the second codon letters.

Nevertheless as one can see the increasing of the amino acids molecular mass is followed mainly with increasing of the molecular mass of the second anticodon (codon) base. The rule related to the base doublets mentioned above (*Table* 1) is much stricter. These rules are probably a consequence of a very curious arithmetic of the code matrix which describes the design of the genetic code shown in this article.

Digitization of the matrix (numbering of amino acids and nitrogen bases sequences ordered by molecular mass) shows its symmetrical pattern, but also marks these symmetries with very unusual values. The singularity of these numbers excpressed with positional notations is in their simmetry by orders. Anyway these findings support earlier data of the arithmetic content of the genetic code. This content can be seen using regular (natural series of numbers 1-20) and irregular sequences of values of nucleon masses of variable parts of molecules (1-130 for amino acids, and 35-75 for nitrogen bases). It is marked with specific *information symbols* in several notations. Whether the code arithmetic is independent, or it is connected somehow with the universality of the code, its amazing stability (including the noise stability) remains to be seen.